\documentclass[usenatbib,a4paper,times,fleqn]{mnras}
\usepackage{graphicx,natbib,times,amssymb,amsmath,pstricks}
\usepackage{aas_macros}
\usepackage{textcomp}

\title[Protoplanetary disc isochrones]{Protoplanetary disc `isochrones' and the evolution of discs in the $\dot{M}-M_{\rm d}$ plane}

\author[Lodato et al.]{{Giuseppe Lodato$^{1}$\thanks{giuseppe.lodato@unimi.it}, 
Chiara E. Scardoni$^{1}$,
Carlo F. Manara$^{2}$ and Leonardo Testi$^{3,4,5}$} \\
$^{1}$Dipartimento di Fisica, Universit\`a Degli Studi di Milano, Via Celoria, 16, Milano, I-20133, Italy \\
$^{2}$Scientific Support Office, Directorate of Science, European Space Research and Technology Centre (ESA/ESTEC), Keplerlaan 1, 2201 AZ Noordwijk, The Netherlands \\
$^{3}$European Southern Observatory, Karl-Schwarzschild-Str. 2, D-85748 Garching, Germany \\
$^{4}$Excellence Cluster Universe, Boltzmannstr. 2, D-85748 Garching, Germany\\
$^{5}$INAF-Osservatorio Astrofisico di Arcetri, Largo E. Fermi 5, I-50125 Firenze, Italy
}
\pagerange{\pageref{firstpage}--\pageref{lastpage}} \pubyear{2016}

\date{}
\begin{document}
\label{firstpage}
\bibliographystyle{mnras}
\maketitle

\begin{abstract}
In this paper, we compare simple viscous diffusion models for the disc evolution with the results of recent surveys of the properties of young protoplanetary discs. We introduce the useful concept of `disc isochrones' in the accretion rate - disc mass plane and explore a set of Montecarlo realization of disc initial conditions. We find that such simple viscous models can provide a remarkable agreement with the available data in the Lupus star forming region, with the key requirement that the average viscous evolutionary timescale of the discs is comparable to the cluster age. Our models produce naturally a correlation between mass accretion rate and disc mass that is shallower than linear, contrary to previous results and in agreement with observations. We also predict that a linear correlation, with a tighter scatter, should be found for more evolved disc populations. Finally, we find that such viscous models can reproduce the observations in the Lupus region only in the assumption that the efficiency of angular momentum transport is a growing function of radius, thus putting interesting constraints on the nature of the microscopic processes that lead to disc accretion.
\end{abstract}


%----------------------------------------------------------------------------------------------------------------
\section{Introduction}

Understanding the evolution of protoplanetary discs is essential in order to  study the process of star and planet formation. Since the pioneering studies of \citet{lyndenbell74}, viscous evolution has been considered as playing a fundamental role  in this context, although  several uncertainties in the mechanisms leading to  disc evolution  still remain. Firstly, the physical origin of disc  'viscosity' is still debated, with the magneto-rotational instability (MRI) \citep{balbus03} as the main candidate in the inner, hotter parts of the disc, and the gravitational instability \citep{kratterlodato16} as the main candidate during the early phases of disc evolution, when the disc is still cold and massive. Recently, non-ideal magneto-hydrodynamical (MHD) simulations have found that a large portion of the disc is actually stable to the MRI, and new models based on the effects of magnetic disc winds have been proposed as a way to remove, rather than redistribute, angular momentum in the disc \citep[e.g.,][]{bai16}. Secondly, it is clear that other, non-viscous mechanisms play a role in disc clearing at later times, such as photo-evaporation \citep{clarke01} and planet-formation. Finally, it is well known  that protoplanetary accretion proceeds `in bursts', such that relatively long period of quiescence, with moderate/low accretion rates (of the order of $\sim 10^{-8-10}M_{\odot}/$yr) are followed by sudden bursts of accretion, during which the accretion rate can rise by a factor of a few \citep[e.g.,][]{costigan14} or even by several orders of magnitude  \citep{audard14}. 

In the context of viscous disc evolution, the analytical self-similar solutions of \citet{lyndenbell74} have often been used to describe to a first approximation the evolution of surface density, mass and accretion rate in a protoplanetary disc. While clearly being a simplification, and certainly not accounting for the above mentioned accretion bursts, such models are able to describe the average behaviour of protoplanetary discs in simple terms. One key parameter in this models is represented by the disc viscosity $\nu$, whose origin, as mentioned above, is still unknown, and it is often parameterized in terms of the \citet{shakura73} prescription whereby
\begin{equation}
\nu=\alpha c_{\rm s}H,
\end{equation}
where $c_{\rm s}$ is the sound speed of the gas in the disc, $H=c_{\rm s}/\Omega$ is the disc thickness, $\Omega$ is the disc angular velocity (assumed to be Keplerian), and $\alpha$ is a dimensionless scale parameter, whose value is variously estimated to be in the range $0.001-0.1$ \citep{hartmann,king07}. 

Viscous accretion discs have been used to reproduce the general observed relationship between mass accretion rates and \emph{stellar} mass. \citet{2005ApJ...625..906M} and \citet{2006A&A...452..245N} completed the first spectroscopic surveys of mass accretion rates in young stellar objects with discs in the Taurus and $\rho$-Ophiuchus star forming regions. % \citet{2006ApJ...648..484H} and 
\citet{2006ApJ...645L..69D} analysed those samples in the framework of viscous accretion discs models and showed that populations of discs derived from a set of initial conditions consistent with observed prestellar cores properties could explain the observed relationship between mass accretion rates and stellar masses and between disc masses and stellar mass. \citet{2006ApJ...645L..69D} also showed that their models predict a very tight correlation 
between the disc mass and the mass accretion rate, which was not observed in the limited set of measurements available then.

Recently, more extensive spectroscopic surveys to measure mass and accretion rates in discs in large and complete samples with an homogeneous method, such as in the Lupus or Chamaeleon~I star forming regions \citep{alcala14,alcala17,manara16a,manara17} offer a way to test such evolutionary models in a statistically coherent way, in particular when combined with mm-interferometry surveys of the same regions \citep{ansdell16,pascucci16}.  A
pioneering analysis of this kind was attempted by \citet{hartmann} in order to give estimates in $\alpha$, combining the measured mass accretion rates with the age of the targets to derive how the former decreased with time. \citet{jones12} have performed an analysis of how well do the simple self-similar models of \citet{lyndenbell74} reproduce the evolution of accretion discs, arguing that in order to reproduce the observed mass, accretion rates and age of a sample of accretion discs, significant deviations from the self-similar models were required.  The analysis of \citet{jones12} was limited by several factors. Their sample was collected from a number of sources, and thus resulted to be highly inhomogeneous, and the uncertainties in the determination of almost all disc observables  was very high. In particular, the age of young stars is notoriously challenging to measure, as it has to rely of pre-main sequence evolutionary tracks, that are very  uncertainties at young ages \citep{2002A&A...382..563B,2010A&A...521A..44B,soderblom14}. Recently, \citet{rosotti17} expanded the analysis by \citet{jones12} to study how several effects, such as the presence of dead zones in the disc, planet formation, and external photoevaporation affect the evolution of discs. Also, \citet{rafikov17} has used similarity solutions to estimate the viscosity based on observed disc mass, accretion rates and disc radii in Lupus. 

In this paper, we perform a similar analysis but with two important differences. Firstly, from the observational point of view, we rely on the homogeneous sample of mass accretion rates and disc masses in the Lupus star forming region obtained by \citet{alcala14,alcala17} and \citet{ansdell16} and already combined and analyzed by \citet{manara16b}. Secondly, we introduce the concept of protoplanetary disc `isochrones', that is the locus of points in the $\dot{M}-M_{\rm d}$ plane occupied by a number of sources assumed to have the same age. Analogously to the case of stellar isochrones, this analysis can test viscous evolutionary disc models and return an estimate of the age of a given cluster of protostars,  thus alleviating one of the uncertainties of previous analyses. 

The paper is organized as follows. In section \ref{sec:isochrones} we introduce the concept of disc isochrones, we recall the self-similar solutions and provide an analytical expression  to evaluate them. In section \ref{sec:montecarlo} we perform some Montecarlo simulations, where we produce a synthetic population of protoplanetary discs and place them in the $\dot{M}-M_{\rm d}$ plane, to check the usefulness of the isochrones. In section \ref{sec:lupus} we apply our modeling to the data obtained by \citet{manara16b} in the Lupus region. In section \ref{sec:conclusions} we draw our conclusions. 

\section{Disc `isochrones'}
\label{sec:isochrones}

\subsection{The self-similar solutions}

The (viscous) evolution of a protoplanetary disc is determined by the function $\Sigma(R,t)$ that describes the surface density of the disc as a function of time $t$ and radius $R$. The simplest case that is often used for protoplanetary discs is represented by the so-called `self-similar' solutions, obtained by \citet{lyndenbell74}, under the assumption that disc viscosity has a simple power-law dependence on radius:
\begin{equation}
\nu=\nu_0(R/R_0)^\gamma,
\end{equation}
where $R_0$ is a scale radius, $\nu_0$ the value of viscosity at $R_0$ and $\gamma$ is a free index. The self-similar solution is:
\begin{equation}
\Sigma(R,t) = \frac{M_0}{2\pi R_0^2}(2-\gamma)\left(\frac{R}{R_0}\right)^{-\gamma} T^{-\eta}\exp\left(-\frac{(R/R_0)^{(2-\gamma)}}{T}\right),
\label{eq:sigmass}
\end{equation}
where $M_0$ is the disc mass at $t=0$, $\eta=(5/2-\gamma)/(2-\gamma)$, $T=1+t/t_{\nu}$ and the viscous time is defined as:
\begin{equation}
t_{\nu}=\frac{R_0^2}{3(2-\gamma)^2\nu_0}. 
\label{eq:tv}
\end{equation}
From Eq.~(\ref{eq:sigmass}) one thus sees that the scale radius $R_0$ represents the initial truncation radius of the disc. The disc truncation radius, according to Eq. (\ref{eq:sigmass}), evolves with time as $R_{\rm out}=R_0T^{1/(2-\gamma)}$.
The disc mass is obtained by simply integrating Eq. (\ref{eq:sigmass}) over radius:
\begin{equation}
M_{\rm d}(t)=M_0T^{(1-\eta)}.
\label{eq:massss}
\end{equation}
Note that in order to have a disc mass that decreases with time, we need to require $\gamma<2$. In this scenario, the only mass that accretes onto the star is the mass lost by viscous spreading by the disc and it is thus given by the opposite of the time derivative of $M_{\rm d}(t)$:
\begin{equation}
\dot{M}=-\frac{\mbox{d}M_{\rm d}}{\mbox{d}t}=(\eta-1)\frac{M_0}{t_{\nu}}T^{-\eta}.
\label{eq:mdotss}
\end{equation}
A frequently used parameter when describing disc evolution is the `disc lifetime', defined as the ratio of disc mass to accretion rate \citep{jones12,rosotti17}:
\begin{equation}
t_{\rm disc}=\frac{M_{\rm d}(t)}{\dot{M}_{\rm acc}(t)}=2(2-\gamma)(t+t_{\nu}).
\label{eq:tdisc}
\end{equation}
It is worth noting that, apart from a constant factor, the disc lifetime $t_{\rm disc}$ is a  measure of the age of the disc if $t\gg t_{\nu}$ and a measure of its viscous time if $t\ll t_{\nu}$. 

\begin{figure}
\centering
\includegraphics[width=0.48\textwidth]{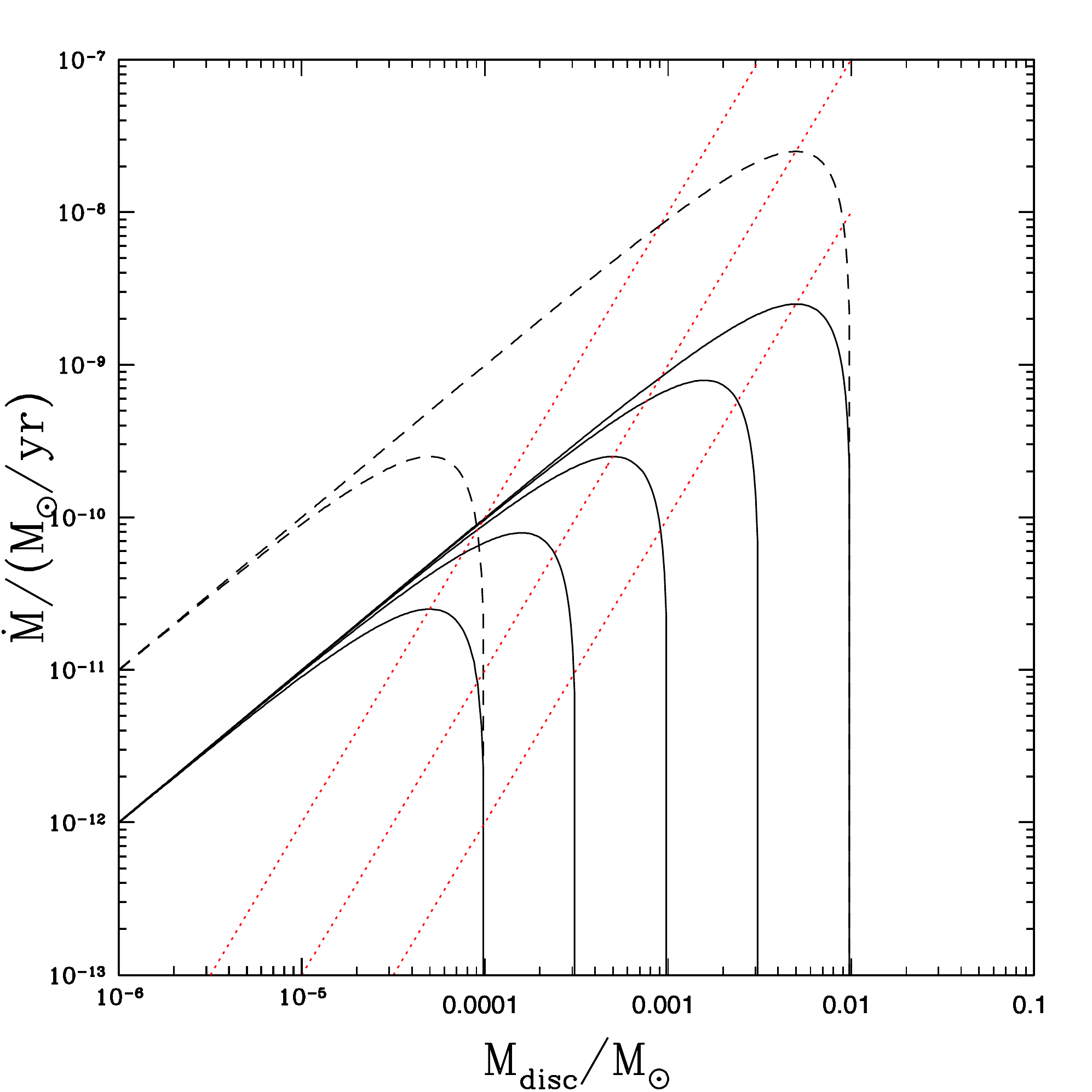}
\caption{Example of protoplanetary disc isochrones, for $\gamma=3/2$. The solid lines show a series of isochrones at an age $t=10^6$ yr, with different initial disc masses $\log(M_0/M_{\odot})=-4$, -3.5, -3, -2.5 and -2, respectively. The dashed lines refer to  $t=10^5$ yrs for $\log(M_0/M_{\odot})=-4$ and -2, respectively. The red dotted lines show instead a few examples of evolutionary tracks, referring to $M_0=10^{-2}M_{\odot}$ and $\log(t_\nu/\mbox{yr})=6$, 5 and 4 (from right to left in the plot).}
\label{fig:isochrone}
\end{figure}

\subsection{Isochrones for the self-similar solution}

We define the isochrone of a population of protoplanetary discs as the locus in the $\dot{M}-M_{\rm d}$ plane of all the discs of the same age $t$ and same initial disc mass $M_0$, defined parameterically as a function of the viscous time $t_{\nu}$. The isochrones for the self-similar solution depends thus on the free parameter $\gamma$. By combining Eqs. (\ref{eq:massss}) and (\ref{eq:mdotss}) it is possible to write the isochrone, for a given initial mass $M_0$, explicitly as:
\begin{equation}
\dot{M}=\frac{M_{\rm d}}{2(2-\gamma)t}\left[1-\left(\frac{M_{\rm d}}{M_0}\right)^{2(2-\gamma)}\right].
\end{equation}
Some example of isochrones, for the case $\gamma=3/2$, are shown in Fig. \ref{fig:isochrone}. The solid lines refer to an age of $t=10^6$ yr, for different $\log(M_0/M_{\odot})=-4$, $-3.5$, $-3$, $-2.5$ and $-2$. The dashed lines refer to  $t=10^5$ yrs for $\log(M_0/M_{\odot})=-4$ and $-2$, respectively. 

The isochrone can be divided into two regions: for low masses and accretion rates (on the left in Fig. \ref{fig:isochrone}) there is a linear relation between $\dot{M}$ and $M_{\rm d}$, corresponding to relatively evolved discs, with $t\gg t_{\nu}$, such that both quantities are simple power-laws with respect to time. On the right side of Fig. \ref{fig:isochrone} we find instead those systems that have not yet had time to evolve and for which the mass is still close to the initial mass $M_0$.  

The disc isochrone should not be confused with the evolutionary track of a single disc with a given $t_{\nu}$, that can be obtained easily by eliminating the age $t$ between equations (\ref{eq:massss}) and (\ref{eq:mdotss}):
\begin{equation}
\label{eq:evtrack}
\dot{M}=\frac{1}{2(2-\gamma)}\frac{M_0}{t_{\nu}}\left(\frac{M_{\rm d}}{M_0}\right)^{5-2\gamma},
\end{equation}
that defines a steeper relation in the $\dot{M}-M_{\rm d}$ plane, that (unlike the isochrone in the self-similar part) depends on the parameter $\gamma$.  A few examples of evolutionary tracks are shown in Fig. \ref{fig:isochrone} as red dotted lines, referring to $M_0=10^{-2}M_{\odot}$ and $\log(t_\nu/\mbox{yr})=6$, 5 and 4 (from right to left in the plot).

Note that the various isochrones evaluated at the same age, but with different initial masses, all tend to the same linear asymptotic relation, whose normalization depends on the isochrone age. Therefore, if we consider an ensemble of discs with different viscous times and initial masses, we would expect to find a tight correlation between mass and accretion rate for low disc masses. On the contrary, for higher disc masses, we would probe those discs that have evolved the less and thus keep "memory" of their initial condition. In this case, the distribution of discs in the plane should maintain the original scatter and thus show a looser correlation. In order to check this behaviour, we have run some Montecarlo realization of initial conditions, that we describe in the next Section.

\section{`Disc population synthesis' models}
\label{sec:montecarlo}

\begin{figure}
\centering
\includegraphics[width=0.48\textwidth]{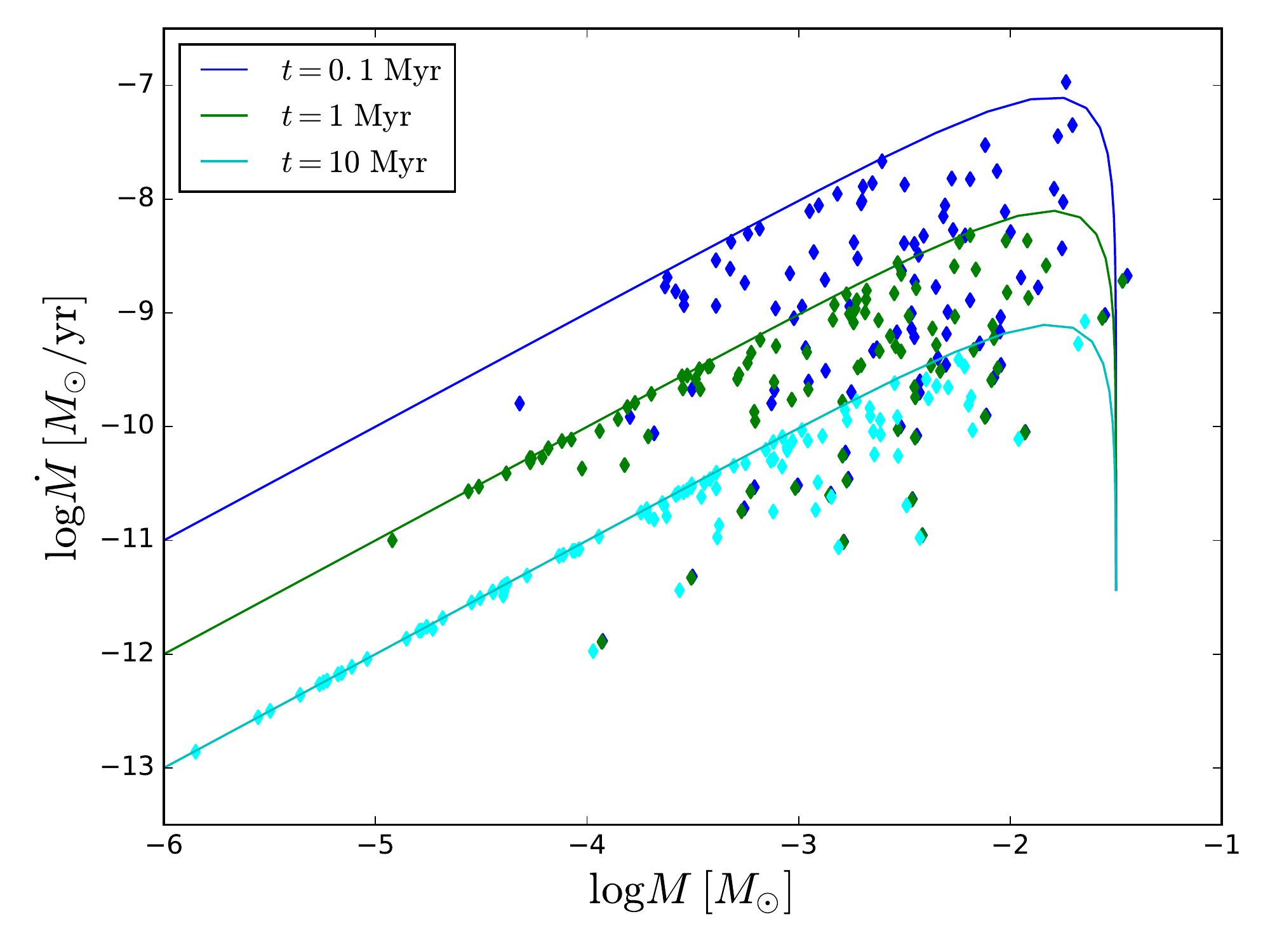}
\caption{Results of a Montecarlo `disc population synthesis' model. A Montecarlo realization of 100 initial conditions for disc evolution, assuming a log-normal distribution in $M_0$ and $t_\nu$ has been produced (see text for details). The plot shows the location of the disc population in the $\dot{M}-M_{\rm d}$ plane for $t=10$ Myr (cyan diamonds), $t=1$ Myr (green diamonds) and $t=0.1$ Myr (blue diamonds). The three lines show the corresponding isochrones (with the same colour scheme), assuming $\log(M_0/M_{\odot})=-1.5$.}
\label{fig:montecarlo}
\end{figure}

We have simulated initial conditions for a sample of  100 protoplanetary discs. In the following, unless otherwise stated, we will always consider the case where $\gamma=3/2$. We have assumed a log-normal distribution of initial masses and viscous times, assuming the following parameters: $\langle\log (t_{\nu}/\mbox{yr})\rangle = 7$, $\sigma_{t_\nu}=1$, $\langle\log(M_0/M_{\odot})\rangle=-2.5$, $\sigma_{M_0}=0.5$, where $\sigma_{t_\nu}$ and $\sigma_{M_0}$ are the Gaussian widths for $\log t_\nu$ and $\log M_0$, respectively.  We then evolved the population up to a fixed age $t$ and plotted the resulting values of disc mass and accretion rates in Fig. \ref{fig:montecarlo}, for three different ages: $t=0.1$ Myr, $t=1$ Myr and $t=10$ Myr, shown with the blue, green and cyan diamonds.  In Fig. \ref{fig:montecarlo} we also plot, for comparison the corresponding isochrones, assuming $\log(M_0/M_{\odot})=-1.5$. This plot confirms the predictions that we made in the previous Section: at $t=10$ Myr most discs have evolved to the self-similar part of the viscous diffusion evolution and indeed display a tight correlation in the $\dot{M}-M_{\rm d}$ plot. At earlier ages, instead, there is progressively  more scatter, especially for high disc masses. 

An important corollary of the fact that the scatter is more pronounced for high disc masses with respect to the lower disc masses (opening in a `fan'-like way below the corresponding isochrone) is that, if one wants to obtain a correlation from the disc population, the resulting correlation will be less steep than linear, a feature that is observed, for example, in the Lupus sample \citep{manara16b}, where $\dot{M}\propto M_{\rm d}^a$, with $a\approx 0.7$, and even when combining the disc masses and mass accretion rates in the Lupus and in the Chamaeleon~I regions \citep{mulders17}. However, due to the limited sensitivity of current surveys, the disc mass range spanned by observations is too little to indicate any mass dependent spread in the correlation.

Additionally, we would expect that, for an evolved sample, where most of the discs have evolved to $t\gg t_\nu$, the scatter in the disc mass - accretion rate plane should be small, while it should progressively increase for younger systems. We are thus left with an interesting possibility of disentangling the degeneracy between age and viscous time inherent from Eq. (\ref{eq:tdisc}): while a measure of the average disc lifetime for a sample of protoplanetary discs gives a measure of the sum $t+t_\nu$, a measure of the scatter in the correlation gives us a measure of $t/t_\nu$.  We will perform such an analysis on the Lupus data in the following Section.

\begin{figure}
\centering
\includegraphics[width=0.48\textwidth]{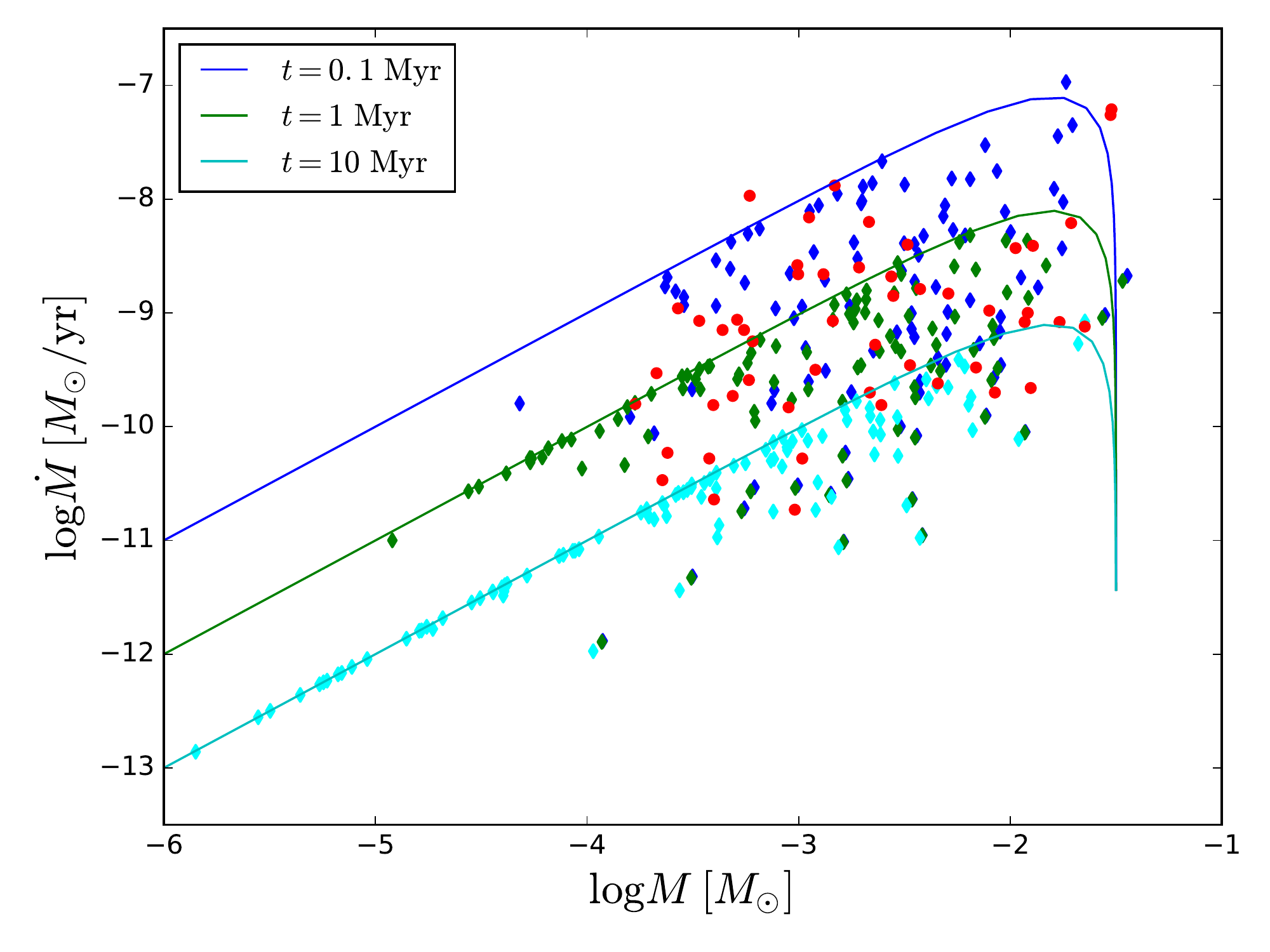}
\caption{Same as Fig. \ref{fig:montecarlo}, but with the data from the Lupus survey \citep{manara16b} overlaid with red circles.}
\label{fig:montecarlo2}
\end{figure}

\begin{figure*}
\centering
\includegraphics[width=0.48\textwidth]{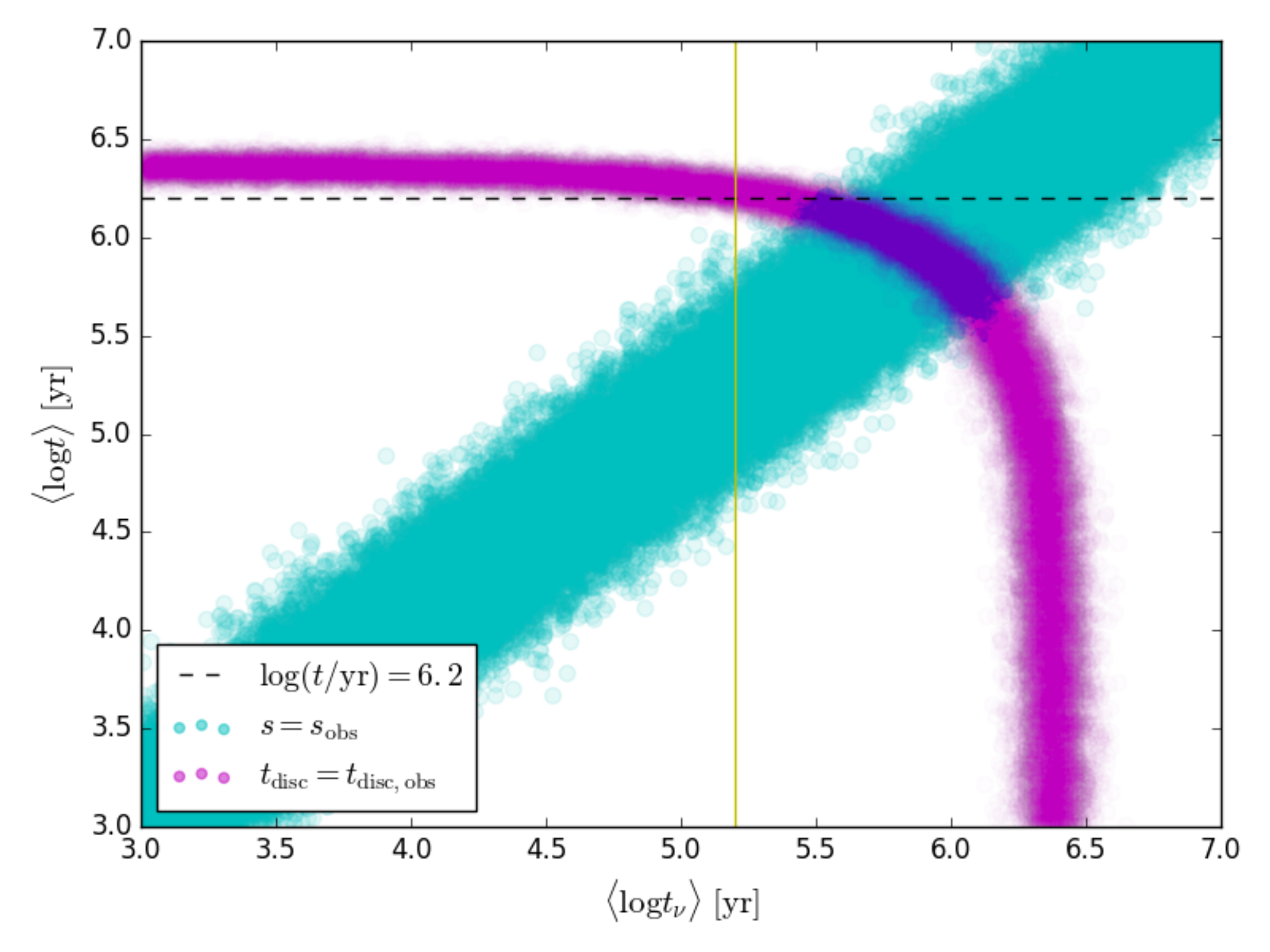}
\includegraphics[width=0.48\textwidth]{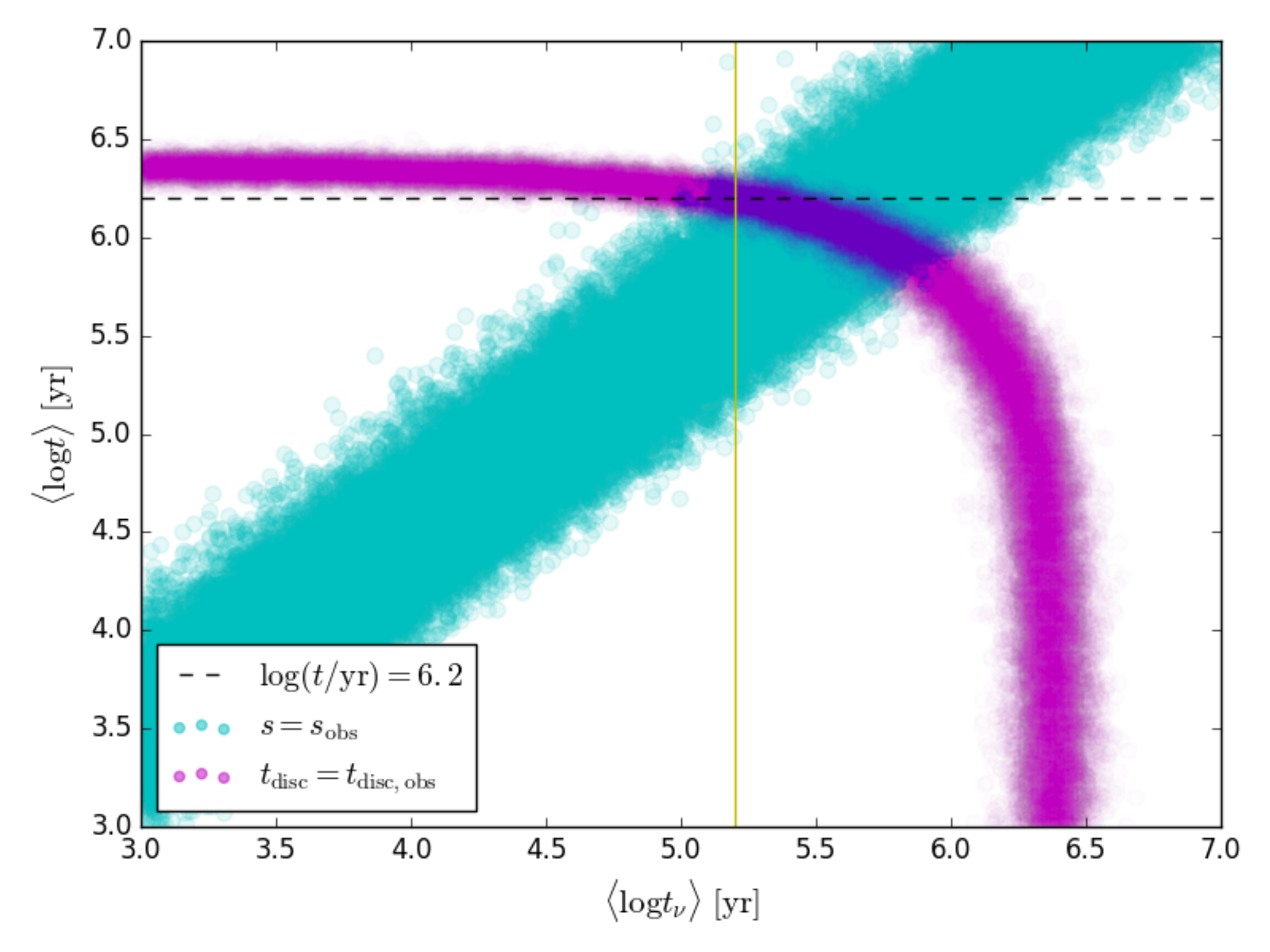}
\caption{Combination of disc population synthesis parameters that match within 10\% the observed value of the average disc lifetime $\langle\log(t_{\rm disc,obs}/\mbox{yr})\rangle = 6.4$ (purple points) and the observed value of the scatter in the $\dot{M}-M_{\rm d}$ plane (cyan points). The left panel refers to $\sigma_{t_\nu}=1$, while the right panel refers to $\sigma_{t_\nu}=1.2$. The horizontal dashed line indicates the known average age in Lupus, while the yellow vertical line indicates the expected viscous time for a disc whose initial truncation radius was $R_0=100$ au, assuming $\alpha=0.1$ and $H/R=0.1$, evaluated at $R=R_0$.}
\label{fig:lupus}
\end{figure*}

\section{An application to the Lupus star forming region}
\label{sec:lupus}

In Fig. \ref{fig:montecarlo2} we plot the same data as in Fig. \ref{fig:montecarlo}, but with the observed data from the Lupus survey \citep{manara16b} overlaid with red circles. The mass accretion rates shown here have been derived measuring the excess emission in the Balmer continuum region from broad-band flux-calibrated VLT/X-Shooter spectra \citep{alcala14,alcala17}, while the disc masses are obtained from the 0.87 mm continuum flux \citep{ansdell16} and converted in total disc mass assuming a gas-to-dust ratio of 100.  Note that the viscous disc models give a prediction for the gas disc masses. We have chosen to use dust masses converted into gas masses using a standard gas-to-dust ratio because accurate measurements of gas masses based on $CO$ emission are debated \citep{miotello16}.  
Here we include in the analysis all the discs detected at mm-wavelengths. We considered the measured values for the mass accretion rate for all these targets, including those where the measured accretion rates are compatible with the chromospheric emission \citep[see discussion by ][]{alcala17,manara17}. The discs that are undetected at mm-wavelengths are excluded, but this does not alter the overall scatter of the data along the best fit \citep[see Fig.~1 of][]{manara16b}.  
It is apparent that the Lupus data show a significant scatter, much larger than the one expected from the models at $t\approx 1$ Myr, which is of the order of the average age of the Lupus stars. In the following we assume the age of this region to be $t_{\rm Lupus}=1.6$ Myrs, with a spread of $\approx 0.3$ dex, based on the estimates by \citet{comeron08} or \citet{alcala17}, ranging from 1 to 3 Myr.  However, it should be kept in mind that Fig. \ref{fig:montecarlo2} has been produced by evolving our simulated disc population to a fixed given age, while in reality, as mentioned above, we know that there is a significant spread in ages as well in the observed sample.

We have thus re-run our disc population synthesis model (using 300 simulated discs) by allowing not only a spread in initial conditions (as above), but also a spread in ages. Additionally, we have also simulated an observational error on the measured masses and accretion rates by adding an additional stochastic spread of  0.1 dex for the final disc masses and 0.45 dex for the final
accretion rates from the model.

We have then run a large number of models, varying the average viscous time $\langle\log t_\nu\rangle$  and the average age $\langle\log t\rangle$, and keeping a fixed Gaussian width $\sigma_{t_\nu}$ equal to either 1 or 1.2 dex (see below) and a given width $\sigma_t$, taken to be equal to the known 0.3 dex dispersion in Lupus ages. 

In order to compare these models to the data, we can define the average disc lifetime for a given population as $t_{\rm disc}=10^{\tau}$ yrs, where $\tau=\langle (\log M-\log{\dot{M}})\rangle$. The scatter of the population is then defined as $s=\langle (\log M-\log{\dot{M}}-\tau)^2\rangle$. We then compute $t_{\rm disc}$ and $s$ for each population (either observed or obtained from our population synthesis models and retain only those models that match with a given precision the observational parameters. We have computed $t_{\rm disc,obs}$ and $s_{\rm obs}$ of the data, obtaining $t_{\rm disc,obs}=2.5$ Myr and $s_{\rm obs}=0.32$. We have then computed $t_{\rm disc}$ and $s$ of each model we have run, and thanks to the comparison between these values and the observed ones, we found the models which match the data.

The results of this modeling are shown in Fig. \ref{fig:lupus}, where we show in purple the combinations of $\langle\log t_\nu\rangle$ and $\langle\log t\rangle$ that match within 10\% (which is the typical uncertainty in the derived parameters) the observed average $t_{\rm disc}$ in Lupus, which is $\langle\log(t_{\rm disc,obs}/\mbox{yr})\rangle = 6.4$, and in cyan the combinations that match the observed scatter. The left panel refers to the case where $\sigma_{t_\nu}=1$, while the right panel refers to $\sigma_{t_\nu}=1.2$. The horizontal dashed line indicates the known average age in Lupus ($\approx 1.6$ Myr), while the yellow vertical line indicates the expected viscous time for a disc whose initial truncation radius was $R_0=100$ au, assuming $\alpha=0.1$ and $H/R=0.1$, evaluated at $R=R_0$. It can be seen that the observed Lupus age falls within our confidence region, marginally for $\sigma_{t_\nu}=1$, and more comfortably for $\sigma_{t_\nu}=1.2$. 

\begin{figure}
\centering
\includegraphics[width=0.48\textwidth]{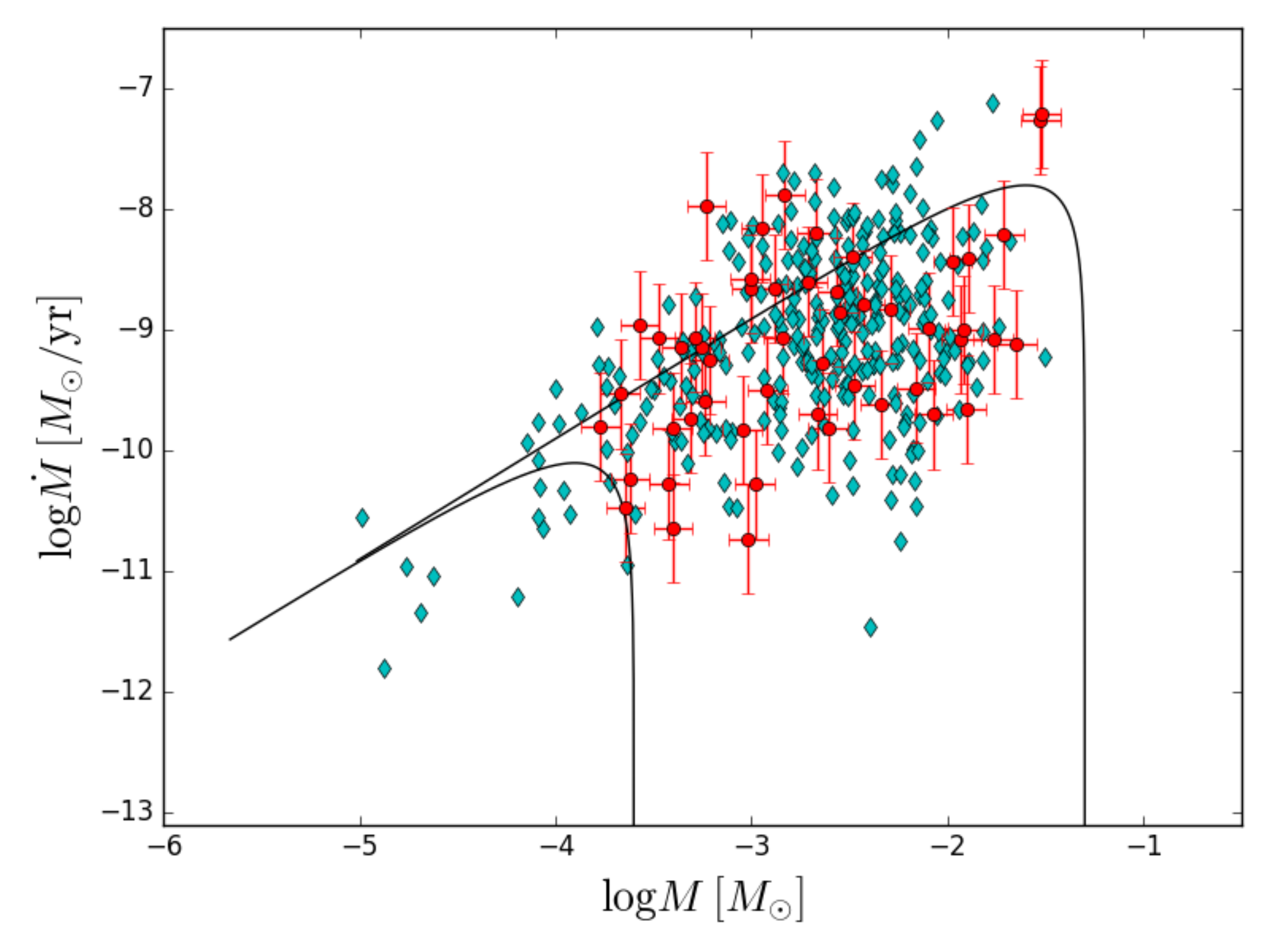}
\caption{Distribution in the $\dot{M}-M_{\rm d}$ plane of our best fit model, for $\sigma_{t_\nu}=1$, with $\langle\log(t/\mbox{yr})\rangle = 5.9$ and $\langle\log(t_\nu/\mbox{yr})\rangle = 5.8$, and further assuming  $\langle\log(M_0/M_{\odot})\rangle=-2.2$ and $\sigma_{M_0}=0.2$. The Lupus data, with error bars, are shown in red, while the diamonds show our disc population synthesis.}
\label{fig:final}
\end{figure}

\begin{figure*}
\centering
\includegraphics[width=0.48\textwidth]{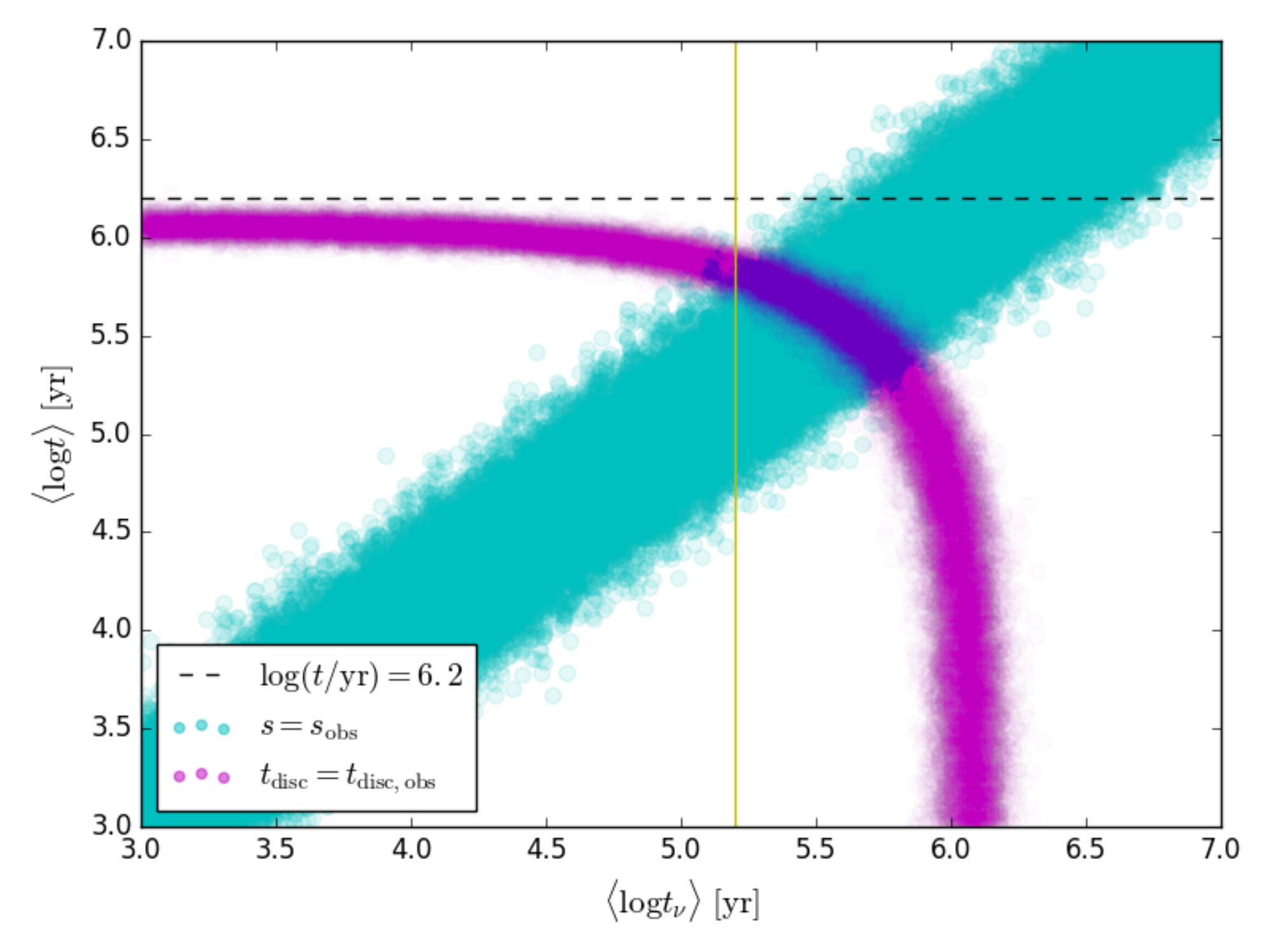}
\includegraphics[width=0.48\textwidth]{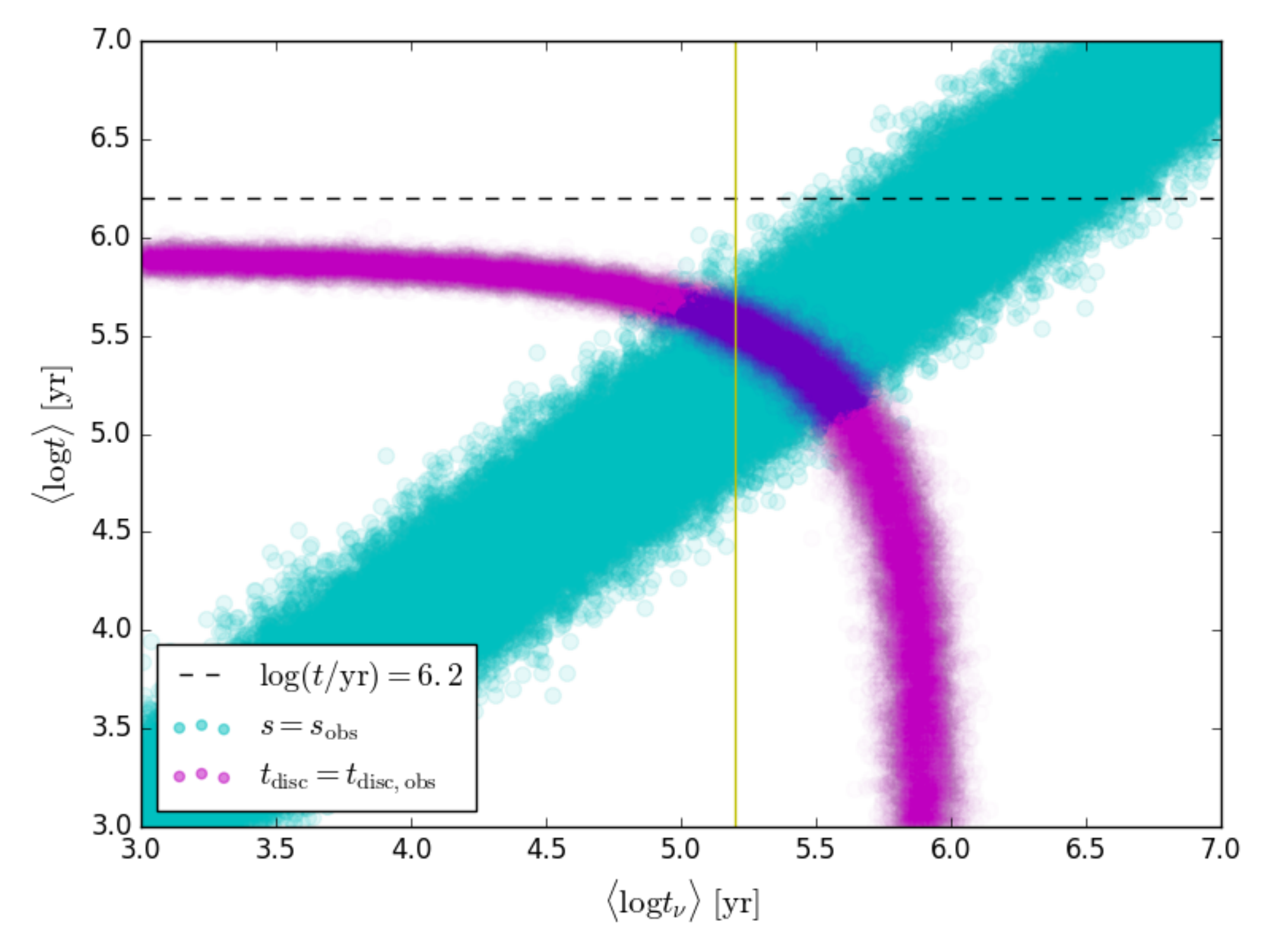}
\caption{Same as Fig. \ref{fig:lupus}, but for $\gamma=1$ (left panel) and $\gamma=0.5$ (right panel).}
\label{fig:b}
\end{figure*}

The interpretation of the purple region in Fig. \ref{fig:lupus} is straightforward. Models on the left horizontal purple branch correspond to $t\gg t_\nu$ and thus the average age must reproduce the observed average disc lifetime (see Eq. (\ref{eq:tdisc})). However, these models provide a scatter that is too small compared to the observed one. On the contrary, the vertical purple line in the right part of the plot correspond to $t\ll t_\nu$, for which it is the viscous time that  reproduces the observed disc lifetime. Such models provide a scatter that is much larger than the observed scatter. In order to provide a match to both the average disc lifetime and the observed scatter, we have to sit in an intermediate class of models, for which the age and viscous times are comparable.  In particular, our confidence regions for the age and viscous time in Lupus, corresponding to the intervals that provide a match to the observed data with the required accuracy, are:
\begin{eqnarray}
\langle\log(t/\mbox{yr})\rangle & = & 5.9\pm 0.3,\\
\langle\log(t_\nu/\mbox{yr})\rangle & = & 5.8\pm 0.4,
\end{eqnarray}
for the $\sigma_{t_\nu}=1$ case, and 
\begin{eqnarray}
\langle\log(t/\mbox{yr})\rangle & = & 6.05\pm 0.25,\\
\langle\log(t_\nu/\mbox{yr})\rangle & = & 5.5\pm 0.5,
\end{eqnarray}
for the $\sigma_{t_\nu}=1.2$ case. 
Note that up to now we did not have to make any assumption on the distribution of initial truncation radii (which are built in our model through the relation between viscous time and truncation radius, Eq. (\ref{eq:tv})).  At the same time, we did not have to define any specific model for the disc viscosity, for example in terms of $\alpha$. 
If we want to translate the average viscous times into average initial truncation radii $R_0$ for the discs,  we have to assume some relationship between viscosity and radius. If we assume an $\alpha$ viscosity, with $\alpha=0.1$ and $H/R=0.1$, evaluated at $R_0$, we obtain $\langle \log(R_0/\mbox{au})\rangle=2.3\pm0.2$, for $\sigma_{t_\nu}=1$, and $\log(R_0/\mbox{au})\rangle=2.1\pm0.2$, for $\sigma_{t_\nu}=1.2$. Clearly these estimates of initial disc radii reflect our choice of $\alpha$ and are expected to be smaller for smaller $\alpha$, since, for a given $t_\nu$, we have that $R_0\propto \alpha^{1/2}$.

In principle, we could have obtained solutions that bracket the observed Lupus age better by allowing an even larger $\sigma_{t_\nu}$, and thus moving the cyan points more to the upper left corner of the plots in Fig. \ref{fig:lupus}, but this would have implied extremely small values for the initial radii of our simulated disc population, with $1\sigma$ values for the radii distribution corresponding to $R_0\approx 15$ au, for the case $\alpha=0.1$. While there is some observational evidence that Class I protostars are surrounded by very compact discs \citep{miotello14}, that might be related to magnetic braking during the main infall phase \citep{hennebelle16}, in order to test whether this is a reasonable value for very young discs, systematic measurements of disc sizes in very young protostars are needed. 

In Fig. \ref{fig:final} we show our  `best fit' model for $\sigma_{t_\nu}=1$, with $\langle\log(t/\mbox{yr})\rangle = 5.9$ and $\langle\log(t_\nu/\mbox{yr})\rangle = 5.8$, and further assuming  $\langle\log(M_0/M_{\odot})\rangle=-2.2$ and $\sigma_{M_0}=0.2$. The Lupus data, with error bars, are shown in red, while the diamonds show our disc population synthesis. It can be seen that our modeling does reproduce the observed population well. 

\section{Discussion}

\subsection{Changing the value of $\gamma$}

In the Sections above, we have always assumed that $\gamma=3/2$. This has the nice property that for this value of $\gamma$:
\begin{equation}
t_{\rm disc}=t+t_\nu.
\end{equation}
Here, we discuss what happens if we allow $\gamma$ to have different values. Before showing the results of the disc population synthesis calculations, we can make some analytical predictions.  In the general case
\begin{equation}
t_{\rm disc}=2(2-\gamma)(t+t_\nu)>2(2-\gamma)t,
\end{equation}
or, put in other terms, the ratio of the observed average age of a given population to the average disc lifetime must satisfy:
\begin{equation}
\frac{\langle t\rangle}{\langle t_{\rm disc}\rangle}<\frac{1}{2(2-\gamma)}.
\end{equation}
This condition has been first pointed out by \citet{rosotti17}.  The above equation can be turned around to give a requirement on $\gamma$ for a given observed age and disc lifetime:
\begin{equation}
\gamma>2-\langle t_{\rm disc}\rangle/2\langle t\rangle.
\label{eq:gamma}
\end{equation}
If we evaluate this for Lupus, for which $\langle t_{\rm disc}\rangle\approx 2.5$ Myr and $\langle t\rangle\approx 1.6$ Myr, we obtain $\gamma\gtrsim 1.2$. We should thus expect that models with smaller values of $\gamma$ cannot be able to match the observed data.

In Fig. \ref{fig:b} we show the result of our modeling, assuming $\gamma=1$ (left panel) and $\gamma=0.5$ (right panel). With these values of $\gamma$ the purple line lies always below the dashed line indicating the age of Lupus, and thus our models will always require an age that is much smaller than the Lupus age,  even assuming the extreme case that $t\gg t_\nu$.

\subsection{Systematic uncertainties}

The analysis above is based on observational data that suffer from often quite strong systematic uncertainties. The most important systematics are in the estimates of disc masses, that are based on dust measurement and thus depend on the assumed dust opacity and dust-to-gas ratio, and in the estimates of stellar ages, for which pre-main-sequence tracks are very uncertain at young ages. Disc mass uncertainties, in turn, affect the measurement of the disc lifetime $t_{\rm disc}\propto M_{\rm d}$. This implies that our conclusions based on Eq. (\ref{eq:gamma}) above concerning the dependence of viscosity on radius can certainly be affected by such uncertainties. 

However, we consider it unlikely that the disc masses are very strongly underestimated in our sample. The disc masses in our sample span a large range but are already relatively close to marginal gravitational stability. This can be seen for example in Fig. 10 of \citet{pascucci16}, that shows the disc to star mass ratio of a sample of discs obtained in the same way as in our sample, but for the Chamaleon region (that shares very similar properties to Lupus in terms of disc mass and accretion rate distributions). Thus, systematic uncertainties in disc masses would probably have the effect of decreasing the value of $t_{\rm disc}$. At the same time, \citet{soderblom14} discuss the uncertainties in stellar ages and conclude that these would go in the direction of increasing the stellar age. Both effect would thus make the constraint implied by Eq. (\ref{eq:gamma}) even more stringent than the $\gamma\gtrsim 1.2$ implied by the data as they stand. 

\subsection{Implications on disc evolution}

In order to reproduce the observed Lupus data, as mentioned above, our models, that are based on very standard viscous accretion models, have to satisfy two less standard requirements: (i) that the average viscous time is a factor of a few larger than previously thought and (ii) that the viscosity (and thus the surface density profile) is a steeper function of radius than previously thought. 

Disc dispersal is though to occur within a few Myrs, based on the fraction of stars with infreared excess in star formation regions of different ages \citep{haisch01,fedele10}. Disc dispersal occurs through a combination of various effects, such as planet formation, but most likely due to the interplay of viscous evolution and photoevaporation \citep{alexander14}, that is not included in our models. Pure viscous models usually result in dispersal timescales that are much longer than observed, so that a reduction in the viscous time (as predicted by us) would go in the direction of alleviating the discrepancy. Clearly, a more complete model including photoevaporation would be needed in order to assess the implications of our conclusions on the fraction of stars with infrared excess at any given time. 

For what concerns the steepness of the viscosity law and of the surface density profile, observational data are not conclusive. The best observations of surface density profile in the gas component is provided by the nearby TW Hya disc \citep{zhang17}. These authors obtain a value of $\gamma=0.9\pm 0.4$ for TW Hya. Now, while their best fit value is lower than our prediction (that, we remind is a prediction for the average population and not for any individual system), our required value of $\gamma\gtrsim 1.2$ lies comfortably within their uncertainties. Another benchmark can be done instead by comparing our value of $\gamma$ with the slope of the Minimum Mass Solar Nebula (MMSN), that is $\gamma=1.5$, perfectly consistent with our estimates.

\section{Conclusions}
\label{sec:conclusions}

In this paper we have introduced two tools that can be useful when interpreting the data obtained from current and future surveys of protoplanetary disc properties in selected star formation regions: the concept of `protoplanetary disc isochrone', for example in the $\dot{M}-M_{\rm d}$ plane, and the  `disc population synthesis' models. 

We have specified our isochrones and disc synthesis models for the simplest class of disc evolutionary models: the self-similar solutions of \citet{lyndenbell74}. By analysing these models we reach the following conclusions:
\begin{enumerate}
\item We expect that a population of discs with similar age should show a tight correlation in the $\dot{M}-M_{\rm d}$ plane only if the age of the system is significantly larger than the average viscous time, and that the scatter should be more prominent for larger disc masses, that are less evolved.
\item As a consequence of the point above, we conclude that viscous disc evolution should result in a slope in the $\dot{M}-M_{\rm d}$ relation that is in general shallower than 1, as observed in the Lupus star forming region \citep{manara16b}, as well as combining the data from this region with those from a similar dataset in the similarly young Chamaleon~I region \citep{mulders17}, unless the discs have ages much larger than their viscous time.
\item Our model, despite being the simplest conceivable viscous evolution model, provides a remarkable fit to the data coming from the Lupus survey.
\item The slope and scatter in the $\dot{M}-M_{\rm d}$ relation are a function of disc age, approaching a slope of 1 and a small scatter for older systems. This prediction of our modeling can be easily tested by looking at the disc populations at different ages. The recent survey of the Chamaleon region \citep{pascucci16} does not reveal any difference with respect to Lupus, given that their ages are very similar. Recently, \citet{ansdell17} have performed an ALMA survey of the older $\sigma-$Ori region, but we still lack sufficiently accurate measurements of accretion rates for this region to properly test our predictions. Additionally, in $\sigma$-Orionis external photoevaporation may alter the disc evolution at late times \citep{rigliaco09}.
\item It is easy to demonstrate analytically that, given the observed ages and disc lifetimes in Lupus, a fit with viscous model can  only be done if the viscosity is a relatively steep function of radius, with $\nu\propto R^\gamma$, and $\gamma\gtrsim 1.2$. This has important implications for what concerns the mechanism of angular momentum redistribution in discs. Indeed, adopting the standard $\alpha$ prescription for viscosity, we have
\begin{equation}
\nu\propto \alpha TR^{3/2}\propto \alpha R,
\end{equation}
where in the last equality we have made the assumption that $T\propto R^{-1/2}$, as usually observed. A slope larger than 1 (as required to reproduce the Lupus data with our modeling) requires that $\alpha$ is a growing function of radius, thus implying an increasing efficiency of angular momentum transport with increasing radius. 
\end{enumerate}

\citet{mulders17} have recently performed a similar analysis, using models with constant $\alpha$ but with a time dependent viscosity \citep{chambers09}.  Their analysis shows, consistent with our findings, that short viscous timescales do not reproduce the observed spread in the mass accretion - disc mass relation. 

Our modeling is obviously simplified and can be further expanded in several directions. From the theoretical point of view, we have just assumed the simplest  viscous evolution for the discs, neglecting several important effects such as planet formation, episodic and variable accretion \citep{audard14}, and photoevaporation. \citet{rosotti17} have considered the effects of both internal and external photoevaporation, of dead zones, and of planet formation, and show that internal photoevaporation, planet formation, or dead zones would lead to a steepening of the isochrone in the low disc mass regime, while external photoevaporation would result in the opposite effect. These effects, when considered, would generate a spread along the relation predicted by viscous evolution. From the observational point of view, the obvious step to do is to test our models on older star forming regions, for which we have provided clear theoretical predictions. 

\section*{Acknowledgments}

We thank an anonymous referee for a thorough reading of our manuscript. We thank the Munich Institute for Astro- and Particle Physics for hospitality during the ``Protoplanetary Disks and Planet Formation and Evolution'' program. We also thank Ilaria Pascucci and Cathie Clarke for interesting discussions. This work was partly supported by the Italian Ministero dell'Istruzione, Universit\`a e Ricerca through the grant Progetti Premiali 2012 – iALMA (CUP C52I13000140001). CFM acknowledges an ESA Research Fellowship.

%, prot. 2010LY5N2T.
\label{lastpage}

\bibliography{lt,biblio,lodato,bibliography}

\begin{thebibliography}{}
\makeatletter
\relax
\def\mn@urlcharsother{\let\do\@makeother \do\$\do\&\do\#\do\^\do\_\do\%\do\~}
\def\mn@doi{\begingroup\mn@urlcharsother \@ifnextchar [ {\mn@doi@}
  {\mn@doi@[]}}
\def\mn@doi@[#1]#2{\def\@tempa{#1}\ifx\@tempa\@empty \href
  {http://dx.doi.org/#2} {doi:#2}\else \href {http://dx.doi.org/#2} {#1}\fi
  \endgroup}
\def\mn@eprint#1#2{\mn@eprint@#1:#2::\@nil}
\def\mn@eprint@arXiv#1{\href {http://arxiv.org/abs/#1} {{\tt arXiv:#1}}}
\def\mn@eprint@dblp#1{\href {http://dblp.uni-trier.de/rec/bibtex/#1.xml}
  {dblp:#1}}
\def\mn@eprint@#1:#2:#3:#4\@nil{\def\@tempa {#1}\def\@tempb {#2}\def\@tempc
  {#3}\ifx \@tempc \@empty \let \@tempc \@tempb \let \@tempb \@tempa \fi \ifx
  \@tempb \@empty \def\@tempb {arXiv}\fi \@ifundefined
  {mn@eprint@\@tempb}{\@tempb:\@tempc}{\expandafter \expandafter \csname
  mn@eprint@\@tempb\endcsname \expandafter{\@tempc}}}

\bibitem[\protect\citeauthoryear{{Alcal{\'a}} et~al.,}{{Alcal{\'a}}
  et~al.}{2014}]{alcala14}
{Alcal{\'a}} J.~M.,  et~al., 2014, \mn@doi [\aap]
  {10.1051/0004-6361/201322254}, \href
  {http://adsabs.harvard.edu/abs/2014A%26A...561A...2A} {561, A2}

\bibitem[\protect\citeauthoryear{{Alcal{\'a}} et~al.,}{{Alcal{\'a}}
  et~al.}{2017}]{alcala17}
{Alcal{\'a}} J.~M.,  et~al., 2017, \mn@doi [\aap]
  {10.1051/0004-6361/201629929}, \href
  {http://adsabs.harvard.edu/abs/2017A%26A...600A..20A} {600, A20}

\bibitem[\protect\citeauthoryear{{Alexander}, {Pascucci}, {Andrews}, {Armitage}
   \& {Cieza}}{{Alexander} et~al.}{2014}]{alexander14}
{Alexander} R.,  {Pascucci} I.,  {Andrews} S.,  {Armitage} P.,   {Cieza} L.,
  2014, \mn@doi [Protostars and Planets VI]
  {10.2458/azu_uapress_9780816531240-ch021}, \href
  {http://adsabs.harvard.edu/abs/2014prpl.conf..475A} {pp 475--496}

\bibitem[\protect\citeauthoryear{{Ansdell} et~al.,}{{Ansdell}
  et~al.}{2016}]{ansdell16}
{Ansdell} M.,  et~al., 2016, \mn@doi [\apj] {10.3847/0004-637X/828/1/46}, \href
  {http://adsabs.harvard.edu/abs/2016ApJ...828...46A} {828, 46}

\bibitem[\protect\citeauthoryear{{Ansdell}, {Williams}, {Manara}, {Miotello},
  {Facchini}, {van der Marel}, {Testi}  \& {van Dishoeck}}{{Ansdell}
  et~al.}{2017}]{ansdell17}
{Ansdell} M.,  {Williams} J.~P.,  {Manara} C.~F.,  {Miotello} A.,  {Facchini}
  S.,  {van der Marel} N.,  {Testi} L.,   {van Dishoeck} E.~F.,  2017, \mn@doi
  [\aj] {10.3847/1538-3881/aa69c0}, \href
  {http://adsabs.harvard.edu/abs/2017AJ....153..240A} {153, 240}

\bibitem[\protect\citeauthoryear{{Audard} et~al.,}{{Audard}
  et~al.}{2014}]{audard14}
{Audard} M.,  et~al., 2014, \mn@doi [Protostars and Planets VI]
  {10.2458/azu_uapress_9780816531240-ch017}, \href
  {http://adsabs.harvard.edu/abs/2014prpl.conf..387A} {pp 387--410}

\bibitem[\protect\citeauthoryear{{Bai}}{{Bai}}{2016}]{bai16}
{Bai} X.-N.,  2016, \mn@doi [\apj] {10.3847/0004-637X/821/2/80}, \href
  {http://adsabs.harvard.edu/abs/2016ApJ...821...80B} {821, 80}

\bibitem[\protect\citeauthoryear{{Balbus}}{{Balbus}}{2003}]{balbus03}
{Balbus} S.~A.,  2003, ARA\&A, \href
  {http://adsabs.harvard.edu/abs/2003ARA%26A..41..555B} {41, 555}

\bibitem[\protect\citeauthoryear{{Baraffe} \& {Chabrier}}{{Baraffe} \&
  {Chabrier}}{2010}]{2010A&A...521A..44B}
{Baraffe} I.,  {Chabrier} G.,  2010, \mn@doi [\aap]
  {10.1051/0004-6361/201014979}, \href
  {http://adsabs.harvard.edu/abs/2010A%26A...521A..44B} {521, A44}

\bibitem[\protect\citeauthoryear{{Baraffe}, {Chabrier}, {Allard}  \&
  {Hauschildt}}{{Baraffe} et~al.}{2002}]{2002A&A...382..563B}
{Baraffe} I.,  {Chabrier} G.,  {Allard} F.,   {Hauschildt} P.~H.,  2002,
  \mn@doi [\aap] {10.1051/0004-6361:20011638}, \href
  {http://adsabs.harvard.edu/abs/2002A%26A...382..563B} {382, 563}

\bibitem[\protect\citeauthoryear{{Chambers}}{{Chambers}}{2009}]{chambers09}
{Chambers} J.~E.,  2009, \mn@doi [\apj] {10.1088/0004-637X/705/2/1206}, \href
  {http://adsabs.harvard.edu/abs/2009ApJ...705.1206C} {705, 1206}

\bibitem[\protect\citeauthoryear{Clarke, Gendrin  \& Sotomayor}{Clarke
  et~al.}{2001}]{clarke01}
Clarke C.~J.,  Gendrin A.,   Sotomayor M.,  2001, MNRAS, 328, 485

\bibitem[\protect\citeauthoryear{{Comer{\'o}n}}{{Comer{\'o}n}}{2008}]{comeron08}
{Comer{\'o}n} F.,  2008, {The Lupus Clouds}.
p.~295

\bibitem[\protect\citeauthoryear{{Costigan}, {Vink}, {Scholz}, {Ray}  \&
  {Testi}}{{Costigan} et~al.}{2014}]{costigan14}
{Costigan} G.,  {Vink} J.~S.,  {Scholz} A.,  {Ray} T.,   {Testi} L.,  2014,
  \mn@doi [\mnras] {10.1093/mnras/stu529}, \href
  {http://adsabs.harvard.edu/abs/2014MNRAS.440.3444C} {440, 3444}

\bibitem[\protect\citeauthoryear{{Dullemond}, {Natta}  \& {Testi}}{{Dullemond}
  et~al.}{2006}]{2006ApJ...645L..69D}
{Dullemond} C.~P.,  {Natta} A.,   {Testi} L.,  2006, \mn@doi [\apjl]
  {10.1086/505744}, \href {http://adsabs.harvard.edu/abs/2006ApJ...645L..69D}
  {645, L69}

\bibitem[\protect\citeauthoryear{{Fedele}, {van den Ancker}, {Henning},
  {Jayawardhana}  \& {Oliveira}}{{Fedele} et~al.}{2010}]{fedele10}
{Fedele} D.,  {van den Ancker} M.~E.,  {Henning} T.,  {Jayawardhana} R.,
  {Oliveira} J.~M.,  2010, \mn@doi [A\&A] {10.1051/0004-6361/200912810}, \href
  {http://adsabs.harvard.edu/abs/2010A%26A...510A..72F} {510, A72}

\bibitem[\protect\citeauthoryear{Haisch, Lada  \& Lada}{Haisch
  et~al.}{2001}]{haisch01}
Haisch K.,  Lada E.,   Lada C.,  2001, ApJ, 552, L153

\bibitem[\protect\citeauthoryear{Hartmann}{Hartmann}{1998}]{hartmann}
Hartmann L.,  1998, Accretion Processes in Star Formation.
Cambridge University Press, Cambridge

\bibitem[\protect\citeauthoryear{{Hennebelle}, {Commer{\c c}on}, {Chabrier}  \&
  {Marchand}}{{Hennebelle} et~al.}{2016}]{hennebelle16}
{Hennebelle} P.,  {Commer{\c c}on} B.,  {Chabrier} G.,   {Marchand} P.,  2016,
  \mn@doi [\apjl] {10.3847/2041-8205/830/1/L8}, \href
  {http://adsabs.harvard.edu/abs/2016ApJ...830L...8H} {830, L8}

\bibitem[\protect\citeauthoryear{{Jones}, {Pringle}  \& {Alexander}}{{Jones}
  et~al.}{2012}]{jones12}
{Jones} M.~G.,  {Pringle} J.~E.,   {Alexander} R.~D.,  2012, MNRAS, 419, 925

\bibitem[\protect\citeauthoryear{{King}, {Pringle}  \& {Livio}}{{King}
  et~al.}{2007}]{king07}
{King} A.~R.,  {Pringle} J.~E.,   {Livio} M.,  2007, MNRAS, \href
  {http://adsabs.harvard.edu/abs/2007MNRAS.376.1740K} {376, 1740}

\bibitem[\protect\citeauthoryear{{Kratter} \& {Lodato}}{{Kratter} \&
  {Lodato}}{2016}]{kratterlodato16}
{Kratter} K.,  {Lodato} G.,  2016, \mn@doi [ARA\&A]
  {10.1146/annurev-astro-081915-023307}, \href
  {http://adsabs.harvard.edu/abs/2016ARA%26A..54..271K} {54, 271}

\bibitem[\protect\citeauthoryear{Lynden-Bell \& Pringle}{Lynden-Bell \&
  Pringle}{1974}]{lyndenbell74}
Lynden-Bell D.,  Pringle J.~E.,  1974, MNRAS, 168, 603

\bibitem[\protect\citeauthoryear{{Manara}, {Fedele}, {Herczeg}  \&
  {Teixeira}}{{Manara} et~al.}{2016a}]{manara16a}
{Manara} C.~F.,  {Fedele} D.,  {Herczeg} G.~J.,   {Teixeira} P.~S.,  2016a,
  \mn@doi [\aap] {10.1051/0004-6361/201527224}, \href
  {http://adsabs.harvard.edu/abs/2016A%26A...585A.136M} {585, A136}

\bibitem[\protect\citeauthoryear{{Manara} et~al.,}{{Manara}
  et~al.}{2016b}]{manara16b}
{Manara} C.~F.,  et~al., 2016b, \mn@doi [\aap] {10.1051/0004-6361/201628549},
  \href {http://adsabs.harvard.edu/abs/2016A%26A...591L...3M} {591, L3}

\bibitem[\protect\citeauthoryear{{Manara} et~al.,}{{Manara}
  et~al.}{2017}]{manara17}
{Manara} C.~F.,  et~al., 2017, preprint, \href
  {http://adsabs.harvard.edu/abs/2017arXiv170402842M} {} (\mn@eprint {arXiv}
  {1704.02842})

\bibitem[\protect\citeauthoryear{{Miotello}, {Testi}, {Lodato}, {Ricci},
  {Rosotti}, {Brooks}, {Maury}  \& {Natta}}{{Miotello}
  et~al.}{2014}]{miotello14}
{Miotello} A.,  {Testi} L.,  {Lodato} G.,  {Ricci} L.,  {Rosotti} G.,  {Brooks}
  K.,  {Maury} A.,   {Natta} A.,  2014, \mn@doi [\aap]
  {10.1051/0004-6361/201322945}, \href
  {http://adsabs.harvard.edu/abs/2014A%26A...567A..32M} {567, A32}

\bibitem[\protect\citeauthoryear{{Miotello}, {van Dishoeck}, {Kama}  \&
  {Bruderer}}{{Miotello} et~al.}{2016}]{miotello16}
{Miotello} A.,  {van Dishoeck} E.~F.,  {Kama} M.,   {Bruderer} S.,  2016,
  \mn@doi [\aap] {10.1051/0004-6361/201628159}, \href
  {http://adsabs.harvard.edu/abs/2016A%26A...594A..85M} {594, A85}

\bibitem[\protect\citeauthoryear{{Mulders}, {Pascucci}, , {Testi}  \&
  {Herczeg}}{{Mulders} et~al.}{2017}]{mulders17}
{Mulders} G.,  {Pascucci} I.~{Manara} C.~F.,   {Testi} L.,   {Herczeg} G.,
  2017, \apj, p. subm.

\bibitem[\protect\citeauthoryear{{Muzerolle}, {Luhman}, {Brice{\~n}o},
  {Hartmann}  \& {Calvet}}{{Muzerolle} et~al.}{2005}]{2005ApJ...625..906M}
{Muzerolle} J.,  {Luhman} K.~L.,  {Brice{\~n}o} C.,  {Hartmann} L.,   {Calvet}
  N.,  2005, \mn@doi [\apj] {10.1086/429483}, \href
  {http://adsabs.harvard.edu/abs/2005ApJ...625..906M} {625, 906}

\bibitem[\protect\citeauthoryear{{Natta}, {Testi}  \& {Randich}}{{Natta}
  et~al.}{2006}]{2006A&A...452..245N}
{Natta} A.,  {Testi} L.,   {Randich} S.,  2006, \mn@doi [\aap]
  {10.1051/0004-6361:20054706}, \href
  {http://adsabs.harvard.edu/abs/2006A%26A...452..245N} {452, 245}

\bibitem[\protect\citeauthoryear{{Pascucci} et~al.,}{{Pascucci}
  et~al.}{2016}]{pascucci16}
{Pascucci} I.,  et~al., 2016, \mn@doi [\apj] {10.3847/0004-637X/831/2/125},
  \href {http://adsabs.harvard.edu/abs/2016ApJ...831..125P} {831, 125}

\bibitem[\protect\citeauthoryear{{Rafikov}}{{Rafikov}}{2017}]{rafikov17}
{Rafikov} R.~R.,  2017, \mn@doi [\apj] {10.3847/1538-4357/aa6249}, \href
  {http://esoads.eso.org/abs/2017ApJ...837..163R} {837, 163}

\bibitem[\protect\citeauthoryear{{Rigliaco}, {Natta}, {Randich}  \&
  {Sacco}}{{Rigliaco} et~al.}{2009}]{rigliaco09}
{Rigliaco} E.,  {Natta} A.,  {Randich} S.,   {Sacco} G.,  2009, \mn@doi [\aap]
  {10.1051/0004-6361/200811535}, \href
  {http://adsabs.harvard.edu/abs/2009A%26A...495L..13R} {495, L13}

\bibitem[\protect\citeauthoryear{{Rosotti}, {Clarke}, {Manara}  \&
  {Facchini}}{{Rosotti} et~al.}{2017}]{rosotti17}
{Rosotti} G.~P.,  {Clarke} C.~J.,  {Manara} C.~F.,   {Facchini} S.,  2017,
  \mn@doi [MNRAS] {10.1093/mnras/stx595}, \href
  {http://adsabs.harvard.edu/abs/2017MNRAS.468.1631R} {468, 1631}

\bibitem[\protect\citeauthoryear{Shakura \& Sunyaev}{Shakura \&
  Sunyaev}{1973}]{shakura73}
Shakura N.~I.,  Sunyaev R.~A.,  1973, A\&A, 24, 337

\bibitem[\protect\citeauthoryear{{Soderblom}, {Hillenbrand}, {Jeffries},
  {Mamajek}  \& {Naylor}}{{Soderblom} et~al.}{2014}]{soderblom14}
{Soderblom} D.~R.,  {Hillenbrand} L.~A.,  {Jeffries} R.~D.,  {Mamajek} E.~E.,
  {Naylor} T.,  2014, \mn@doi [Protostars and Planets VI]
  {10.2458/azu_uapress_9780816531240-ch010}, \href
  {http://adsabs.harvard.edu/abs/2014prpl.conf..219S} {pp 219--241}

\bibitem[\protect\citeauthoryear{{Zhang}, {Bergin}, {Blake}, {Cleeves}  \&
  {Schwarz}}{{Zhang} et~al.}{2017}]{zhang17}
{Zhang} K.,  {Bergin} E.~A.,  {Blake} G.~A.,  {Cleeves} L.~I.,   {Schwarz}
  K.~R.,  2017, \mn@doi [Nature Astronomy] {10.1038/s41550-017-0130}, \href
  {http://esoads.eso.org/abs/2017NatAs...1E.130Z} {1, 0130}

\makeatother
\end{thebibliography}

\end{document}